\documentclass{ws-ijmpe}
\usepackage[super,compress]{cite}
\begin{document}

\markboth{I. Filikhin, A. Karoui, and B. Vlahovic}{Effective Mass of $\alpha$-Cluster in $^{12}$C Nucleus}

\catchline{}{}{}{}{}

\title{Effective Mass of $\alpha$-Cluster in $^{12}$C Nucleus}

\author{I. Filikhin, A. Karoui, and B. Vlahovic}

\address{CREST, North Carolina Central University, Durham,  NC 27707, USA\\
ifilikhin@nccu.edu}

\maketitle

\begin{history}
\received{Day Month Year}
\revised{Day Month Year}
\end{history}

\begin{abstract}
Based on the effective-mass concept, we perform the Faddeev calculations for a low-lying spectrum of 3$\alpha$ states in $^{12}$C nucleus.
A three-body potential is used to describe the known breaking of the 3$\alpha$-cluster structure in the nucleus. We show that the contribution of the three-body potential to the Hamiltonian can be compensated by increasing/decreasing the $\alpha$-particle free mass. The effective-mass values are adjusted so that to reproduce the experimental data for the $^{12}$C nucleus. The energy dependence of the effective mass and the correlation to a three-body potential are discussed. We show that the coupling between the $0^+$ ($2^+$) levels forms a specific picture of anti-crossing on the energy/effective-mass plane.
\end{abstract}


\ccode{PACS numbers: 21.60.G, 71.18.+y, 11.80.Jy}


\section{\label{sec:intro} Introduction}

The effective-mass approach for three-particle systems was proposed in Ref. \cite{FSV21} based on the effect of a mass-energy compensation that takes place for a three-body Hamiltonian.
This compensation manifests the mass-energy equivalence in bound nuclear systems.
It was shown in Ref. \cite{FSV21} that a decrease in kinetic energy can be compensated by an increase in the attractive contribution of potential energy. The corresponding calculations for $^3$H and $^3$He nuclei have been performed using the Faddeev method in the configuration space.
The nucleon effective mass was defined to compensate for a contribution of
three-body nuclear potential. The potential is required
to reproduce 3$N$ experimental data using phenomenological nucleon-nucleon potentials.
The simulations have shown that the results of this efficient model can be comparable to those based on sequential consideration of the three-body force (3BF).
The evaluations for kinetic energy and pair-potential terms of the Hamiltonian have been compared to ones of different authors \cite{KKF, C85,Nogga,KVR} for $^3$H nucleus applying the Argonne AV14 potential and different three-body potentials. It was found that the nucleon effective mass have slightly increased from the free-mass value that the ratio $m^*/m$ of effective mass $m^*$ and free mass $m$ is 1.017. This relatively small correction reflects the contribution of a three-body force in the 3$N$ Hamiltonian.

We have to note that the idea of effective mass is widely applied in solid-state physics \cite{K}. For example, in Ref. \cite{FSV06}, the effective mass of an electron in InAs is 0.024$m_0$, while  in GaAs is 0.064$m_0$, where $m_0$ is the mass of a free electron. In nuclear physics and astrophysics, the effective mass method is used in the study of neutron stars \cite{em} within the more general one-particle method of many-body physics \cite{C}. According to the consideration, an individual particle (nucleon) is in a complex interaction with the environment (other nucleons in nuclei), which affects the particle dynamics \cite{La,B}, and the one-particle potential is considered to be energy dependent \cite{MB}.

Applying the effective mass method, we consider a 3$\alpha$-cluster system simulating low-lying spectral states of the $^{12}$C nucleus.
The $\alpha$-particle model has been proposed almost a century ago (see, for example, \cite{W}) based on the molecular properties of low-lying $^{12}$C states. This molecular aspect has been studied
within the framework of several variational approaches \cite{BI}.
In Refs. \cite{H,FT,VCD,PC,Ka,PAC,FJ,OK,POCM,Hiya,f,Fujiwara,FKKM,KK,FSV2005,KK1,I10,H13}, the cluster approach has been used to treat the 3$\alpha$ problem with
different inter-cluster potentials, phenomenologically obtained.
However, the cluster model has failed in
a detailed description of the $^{12}$C spectrum. The main problem is that this approach
does not reproduce the ground state energy of the
$^{12}$C nucleus taking into account only $\alpha$-$\alpha$ potential.
To reach an admissible description of experimental data via cluster calculations, one needs to add a three-body potential \cite{Hiya,FJ,f,FKKM}. However, the potential would be different for different spectral levels of the 3$\alpha$ system.
Moreover, a three-body potential can be repulsive \cite{OK}, \cite{Hiya,KK} or attractive one \cite{FJ,I10}. That depends on the choosing of a two-body potential model.
For example, a repulsive 3$\alpha$ potential
was phenomenologically introduced in Ref. \cite{KK1} with the dependence on the total angular momentum $ J$=$0^+$, $2^+$, and $4^+$. The potential has the Gaussian form with the depths
$V(0^+)$=31.7~MeV, $V(2^+)$=63.0~MeV, and $V(4^+)$=150.0~MeV \cite{H13}, respectively.
An attractive three-body potential has been proposed in Ref. \cite{FSV2005}. In corresponding 3$\alpha$-cluster model, the phenomenological pair potential having $s$, $d$, and $g$ partial components was used from Ref. \cite{AB}.

In the present work, we implement the cluster model from Ref. \cite{FSV2005}. However, the current model does not include the three-body potential.
We introduce an $\alpha$-cluster effective mass that effectively substitutes the contribution of the three-body potential. In our calculations, an adjustment of effective-mass values has to reproduce the experimental energy for considering levels. The effective mass is assumed to be a function of energy and orbital momentum of the 3$\alpha$ state.
To solve the corresponding three-body problem, we apply the method of the Faddeev equations in configuration space \cite{MerkFad}. That allows us to take into account the Coulomb interaction in a rigorous manner \cite{FSV2005}.

\section{Model}
\label{sec:form}

\subsection{Mass-energy compensation: $3N$ system}
\label{sec:AV14}
In this section, we consider the $AAA$ model for the $^3$H nucleus,  where the masses of protons and neutrons are the same, and the particles in the system are identical. The averaged nucleon mass $m$ was chosen as in Ref. \cite{Nogga} as $m_N$=938.9 MeV (or $\hbar ^2/m_N$ = 41.473 MeV fm$^2$).
Our calculations for the ground state were performed
using the charge-independent AV14 potential \cite{AV14}.
The calculated value for the $^3$H binding energy is equal to -7.684~MeV.
Without three-body potential, the 3$N$ system is weakly bound, and the difference between our AV14 results and experimental values
$\Delta E=E^{Cal.}-E_{exp.}$ is 0.798~MeV.
One can find many examples of corresponding calculations in the literature \cite{C85,Nogga,KVR}
given without and with a three-body potential.
The three-body Hamiltonian reads
\begin{equation}
\label{eq:3}
H=H_0+V_{2bf}+V_{3bf},
\end{equation}
where $H_0$ is three-body kinetic operator, $V_{2bf}$ ($V_{3bf}$) is two-body (three-body) interactions. Here, we assume that the three-body force term is a perturbation. Other factors of the nucleon-nucleon interaction like the charge symmetry breaking or electro-magnetics terms (see Ref. \cite{Nogga1}) are not considered in this model.
Following Ref. \cite{FSV21}, we define the effective nucleon mass $m^*$ in a nucleus. The effective mass can be obtained as a correction to the averaged free nucleon mass to reproduce the experimental value of the ground state energy.
The possibility of the procedure was shown in Ref. \cite{FSV21} due to the
the mass-energy compensation effect takes a place for the three-body Hamiltonian.
To explain this effect, we consider the Hamiltonian written in following form:
\begin{equation} \label{eq:4}
H=\beta H_0+\alpha V_{2bf},
\end{equation}
where $\beta, \alpha> 0$.
We can change the mass and the depth of potential simultaneously by using parameters $\alpha$ and $\beta$ defining the scaling for the nucleon mass by the relation $m^*=m_N/\beta$.
The corresponding energy changes are oppositely directed and can be compensated by each other.
Similar compensation is possible for a three-body force as a part of the potential term.

\begin{figure}[ht]
\begin{center}
\includegraphics[width=20pc]{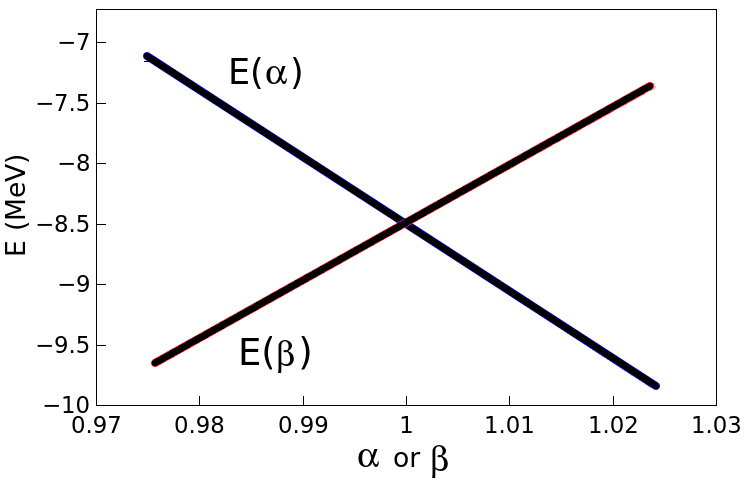}
\end{center}
\caption{ The ground state energy $E$ of $^3$H as a function of the parameter $\alpha$ (the line $E(\alpha$)) or $\beta$ (the line $E(\beta$)) calculated with
the AV14 $NN$ potential and corresponding to the Hamiltonian $H=\beta H_0+\alpha V$, where $\beta, \alpha> 0$, and $m \to m/\beta$ in the $H_0$.
The linear fits for the calculated values are shown. The effective mass of nucleon $m^*_N/m_N$ is equal to 1.017 for $\alpha =\beta=1$.
} \label{fig:5}
\end{figure}

Fig. \ref{fig:5} illustrates the calculations for the ground state using the Hamiltonian (\ref{eq:4}).
Increasing the nucleon mass leads to an increase in the binding energy of the three-body system. The energy $E(\alpha ,\beta )$ is a linear function near the cross point $\alpha=\beta =1$.
The compensation has been interpreted in Ref. \cite{FSV21} as a manifestation of the mass and energy equivalence of  applied to three-nucleon system.

\begin{figure}[ht]
\begin{center}
\includegraphics[width=21pc]{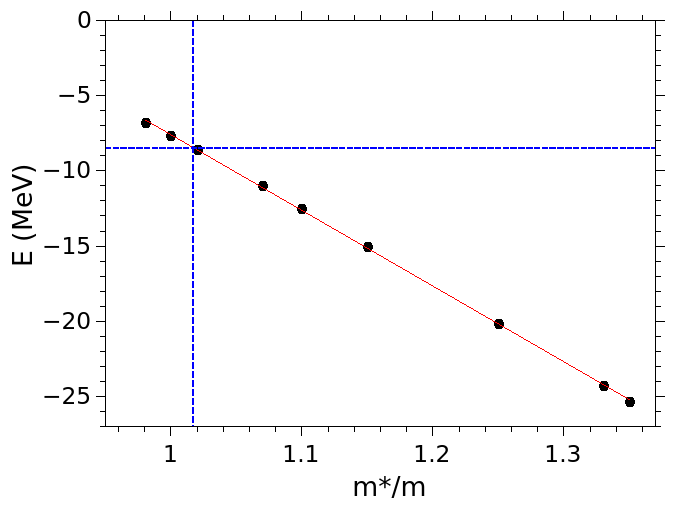}
\end{center}
\caption{ The ground state energy $E$ of $^3$H as a function of the effective mass $m^*/m$. The solid circles are the results of the calculations (the AV14 $NN$ potential). The fine red line corresponds
to the linear fit for the calculated values. The dashed lines correspond to the effective mass of nucleon $m^*/m$ about 1.017 and the experimental value for the ground state energy about -8.48 MeV.
} \label{fig:2}
\end{figure}

The dependence of the $^3$H ground state energy on the effective mass $m^*/m$ could be applied to an evaluate the nucleon effective mass.
We calculate the energy using the AV14 $NN$ potential. 
The results of the calculations fitted by linear functions are presented in Fig. \ref{fig:2}. One can obtain the same linear dependence using the results published in Ref. \cite{W21} for the $NN$ CB-Bonn potential scaled by a parameter.
  Alternatively, such energy dependence on the potential scaling one can repeat by nucleon mass scaling \cite{FSV21}.
In this way,  the effective mass of nucleon $m^*/m$ is about 1.017. That corresponds to the experimental value for the ground state energy of about -8.48 MeV.
In Fig. \ref{fig:2}, we show this correspondence by the vertical and horizontal dashed lines.

It has to note that the result of the calculation for the 3$N$ wave function within the proposed effective approach is comparable with the results of 3$N$ realistic models from the literature.

\subsection{$\alpha $-cluster model for $^{12}$C:  effective-mass extension}
\label{sec:em1}
The $\alpha$-cluster model describes the $^{12}$C nucleus as a three-body system of $\alpha $-particles. The particles interact with pair and three-body nuclear potentials and the Coulomb force. In the present work, we use the formalism of Ref. \cite{FSV2005} based on the Faddeev equations in configuration space. 
Also, we apply phenomenological Ali-Bodmer (AB) $\alpha$-$\alpha $ potential \cite{AB} having different $s$, $d$ and $g$ partial components which  are written as  two-range Gaussian:
$$
V(x)= V_1\exp (-
\frac{x^2}{\beta_1^2}) + V_2\exp (-\frac
{x^2} {\beta_2^2}) 
$$
where the strength and range parameters are
$V_1 $=500~MeV, $ \beta_1 $=1.43~fm, $V_2 $=-130~ MeV, $ \beta_2 $=2.11~fm for $s$-wave;
$V_1 $=320~ MeV, $ \beta_1 $=1.43~fm, $V_2 $=-130~ MeV, $ \beta_2 $=2.11~fm for $d$-wave;
$V_1 $=0, $V_2 $=-130~ MeV, $ \beta_2 $=2.11~fm for $g$-wave.
The attractive $s$-wave component of the potential simulates the Pauli blocking.

The starting point of our approach is related to the kinetic, 2-body, and 3-body potentials terms of the 3-$\alpha$ Hamiltonian. The 2-body potential must reproduce two-body experimental data. Three-body potential can be found by 3-$\alpha$ data. The Ali-Bodmer-type potentials satisfy this request\cite{FSV2005}. The potential has simple coordinate dependence with a repulsive core in $s$ partial wave. The combination AB +3$\alpha$ potentials obtained in Ref. \cite{FSV2005} describes the low-lying $^{12}$C spectrum. The dependence of the 3-$\alpha$ potentials on total orbital momentum was detected. Thus, an additional adjustment for the 3BF potential will reproduce the spectrum with good accuracy. Note here that our goal is to present the general scheme of the effective mass definition for the three-body system. The main assumption of the approach is that a three-body potential acts in the system. Our choice for the pair potential is based on the model proposed in Ref. \cite{FSV2005} by two coauthors of the presented work.

The angular momentum configurations \{$(l,\lambda)$\}, where ${\bf J}={\bf l}+{\bf {\lambda}}$ is total orbital momentum, ${\bf l}$ is momentum
of a pair of particles, ${\bf {\lambda}}$ is the relative momentum of third particle.
One may choose the quantum number sets\cite{FSV2005}  $(0, 0)(2, 2)(4, 4)$ for the
$0^+$ band and $(2, 0)(0, 2)(2, 2)(2, 4)(4, 2)(4, 4)$ for the $2^+$ band.
The $(4, 0)(0, 4)(2, 2)(2, 4)(4, 2)(4, 4)$ configuration
of angular momenta for the $4^+$ states and the $(0, 1)(2, 1)(2, 3)(4, 3)(4, 5)$ configuration for
$1^-$ state.

The known breaking of cluster structure is described by a three-body potential which acts at short distances \cite{FKM05}.
According to the effective-mass concept presented in the previous section, the contribution of the three-particle potential can be compensated by a variation of free mass of $\alpha$-particles, $m_\alpha$. One achieves compensation when the calculated energy is equal to the experimental value. The effective mass $m^*_\alpha$ can be smaller or larger than the free mass depending on the energy level. Obviously, there is a correlation between depth of three-body potential and effective mass value. 

It may be noted that one of the reasons \cite{SK,MF18} for the violation of the cluster structure of $^{12}$C is the significant mixing of the $p_{3/2}$ subshell closure component\cite{SK} to the pure   3$\alpha$-cluster wave function. This mixing generates additional terms in the Hamiltonian. We consider the additional terms as a form of three-body force. In Ref.\cite{SK}, such breaking of the $\alpha$-cluster structure affects all considered states of the $0^+$ and $2^+$ bands like a three-body force. Particularly, in Ref.\cite{SK},  a coupling between the $0^+$  states was discussed. Below we present the evidence for such coupling, obtained in the framework of the effective-mass approach.

The attractive
three-body potential $V_3(\rho)$ having the Gaussian form was defined in Ref. \cite{FSV2005}:
\begin {equation}
V_3(\rho) = V \exp(-\frac{\rho^2}{2\beta^2}),
\label{v_3}
\end {equation}
\noindent where $ \frac{1}{2} \rho^2 =\sum_{i=1}^{i=3} {\bf r}_i^2 $ and $\bf {r}_i$ is
the position vector of ${i}$-th $\alpha$ particle relative to the center of mass of the
system, $V$=-31.935483~MeV, $\beta $ = 3.315~fm.
We use this potential as an auxiliary one for  calculations of resonances. It should be emphasized that the potential (\ref{v_3}) does not affect the two-particle threshold of three-particle systems \cite{FSV2005}.

\section{Numerical results}
\label{sec:res1}
Based on the effective-mass approach, we calculate the low-lying spectrum of a 3$\alpha$-cluster system using the $\alpha$-$\alpha$ Ali-Bodmer potential of the "d" version \cite{AB}. The results of our calculations are presented in Table. \ref{t8}. Here, the effective mass $m_{\alpha}^*$ is adjusted to reproduce the experimental energy of the first two levels of the $0^+$, $2^+$, $4^+$ bands, and $3_1^-$ and $1_1^-$ levels.
The bound states are the states $0^+_1$ and $2^+_1$, and the remaining states are resonances.
In order to calculate the energy of resonances, we used the analog \cite{FSV11} of the method of continuation in coupled constant \cite{KK77,KKH}.
In our study, the effective mass was considered as a coupled constant. To estimate the energy of the resonances, we calculate a set of bound state energies obtained with different effective-mass values.

This evaluation procedure is shown in Fig. \ref{fig:2} for the $^3$H bound state. The simple linear interpolation to the positive energy region allows us to evaluate effective mass by correspondence to the experimental value for energy.
\begin{table}[!b]
\caption{The effective mass $m^*_\alpha/m_\alpha$ with relation to the low-lying levels of the $^{12}$C.
The calculated energies for each level is shown.
The energy is measured from the three-body  $\alpha+\alpha+\alpha$ threshold.
}
\label{t8}
\begin{center}
{\begin{tabular}{lcccccccc} \hline\noalign{\smallskip}
$^{12}$C & $0^+_1$ & $0^+_2$ & $2^+_1$ & $2^+_2$& $4^+_1$ & $4^+_2$ & $3^-_1$ & $1^-_1$\\ \noalign{\smallskip}\hline\noalign{\smallskip}
$E$ (MeV)&-7.275 & 0.286 &-2.93 & 2.25 &    6.05& 6.81& 2.4& 3.6\\
$m^*_\alpha/m_\alpha$&1.290 & 1.176 & 1.15 &1.11 & 0.92  & 1.17& 1.05& 1.19\\
\noalign{\smallskip}\hline
\end{tabular} }
\end{center}
\vskip -0.3cm
\end{table}

\begin{figure}[t]
\begin{center}
\includegraphics[width=19pc]{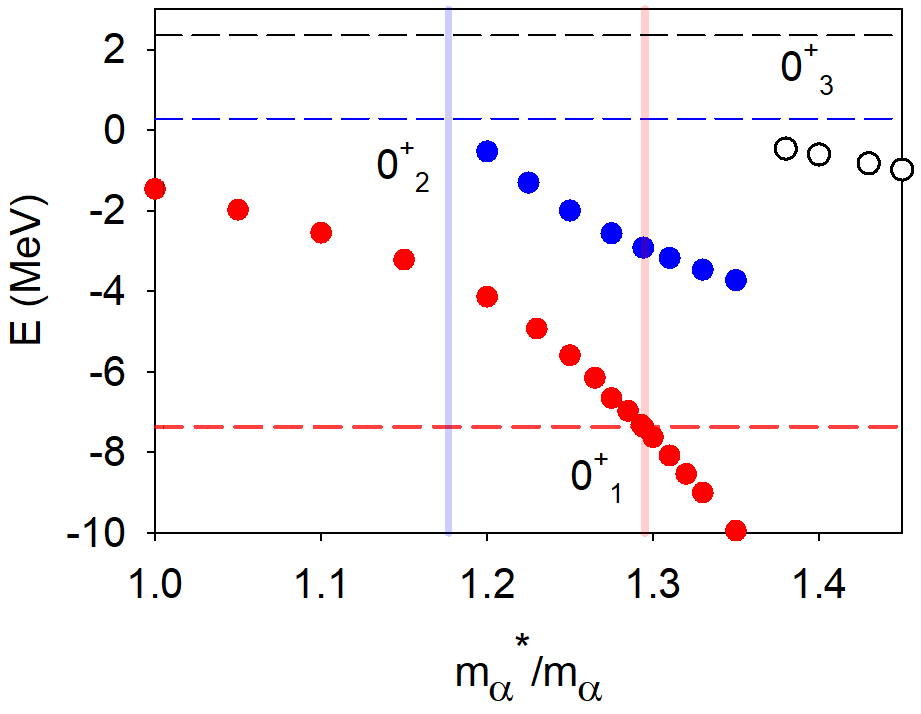}
\includegraphics[width=19pc]{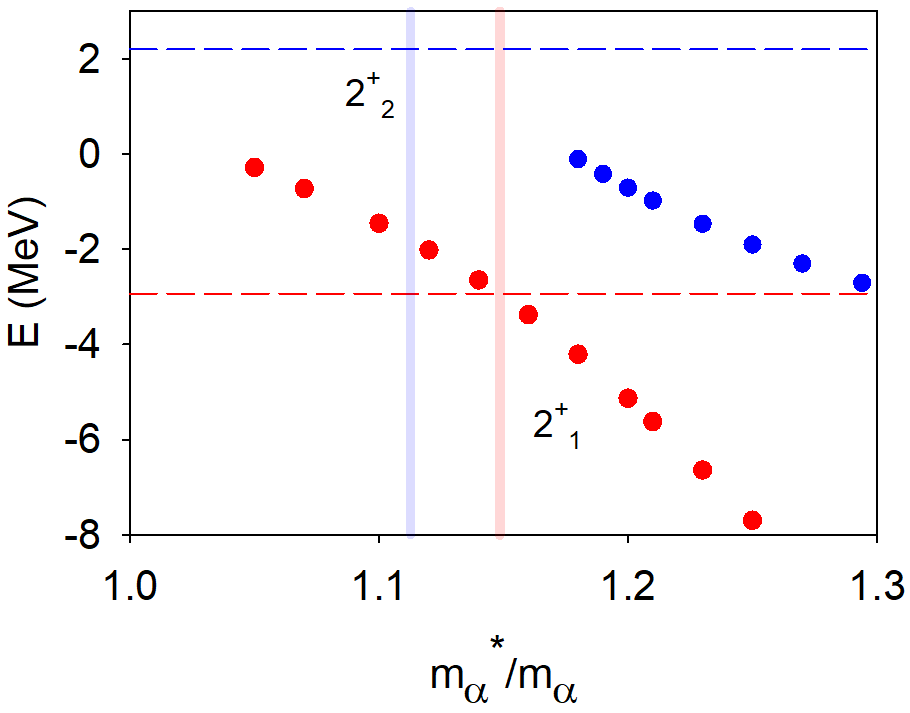}
\end{center}
\caption{ \label{fig:7}
The energy $E$ of the states $J^+_1$ (red solid circles) and $J^+_2$ (blue solid circles), $J=0,2$ and $0^+_3$ (open circles) of $3\alpha$-cluster system along of the $\alpha$-particle effective mass $m_\alpha^*/m_\alpha$. The left panel (right panel) corresponds to
$0^+$ ($2^+$ ) band. The experimental data are shown by the dashed horizontal lines. The vertical lines represent the corresponding values of the effective masses.
The energy is measured from a three-body $\alpha+\alpha+\alpha$ threshold.
}
\end{figure}

In the case of the 3$\alpha$-cluster system, the binding energy and effective mass correspondence can be more complicated.
The results of our calculations for the first levels of the $0^+$ and $2^+$ bands are resented in Fig. \ref{fig:7}. One can see that the energy/effective-mass dependences are non-linear.
We applied a polynomial function to interpolate and extrapolate to negative and positive energies. The physical reason for the non-linear behavior is the coupling between the levels.
Note that the effective mass adjusted by the 0$^+_1$ ground state energy ($m_{\alpha}^*/m_{\alpha}$=1.29) does not lead to the experimental energy for the $0^+_2$ state. This situation is shown in Fig. \ref{fig:7} (Upper panel). To obtain the experimental $0^+_2$ value, one must reduce the effective mass adjusted for the 0$^+_1$ energy to 1.176.
The same situation exists for the $2^+$ states: the effective mass of the $2^+_2$ state is less than one for the $2^+_1$ state.
We explain this statement in Fig. \ref{fig:7} (Lower panel).

The effective mass and three-body potential depth are correlated due to the initial assumption of our approach. Thus, one can assume a dependence of effective-mass value on total orbital momentum. In particular, the effect of the 3$\alpha$ potential in the $2^+$ states must be smaller than one in the $0^+$ states \cite{H13}. For example,
the three-body potential proposed in Ref. \cite{FSV2005} reproduces well low-lying $0^+$ states of the nucleus. However, the same potential applied to the calculations of bound $2^+_1$ and the resonance $2^+_2$ states makes overbounds for the energies of the levels.
An additional adjustment for the parameters of 3BF is needed for 2$^+$ states.
Thus, the hypothesis of the dependence of 3BF on the total orbital momentum can be considered a probable one.

\begin{figure}[!t]
\begin{center}
\includegraphics[width=20pc]{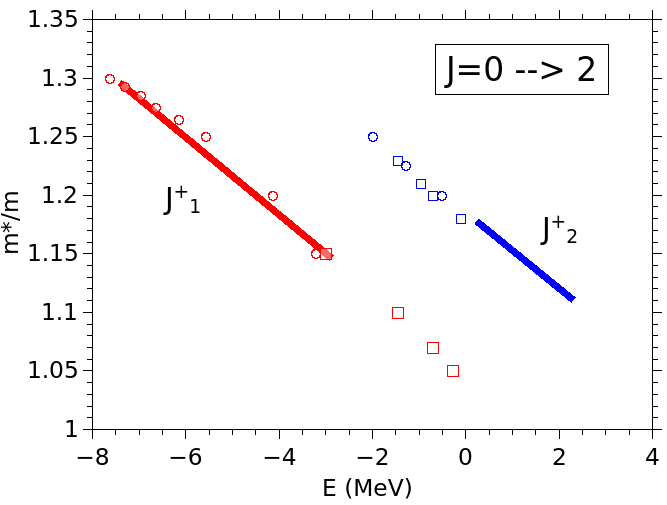}
\end{center}
\caption{ The effective mass ($m^*/m$) (solid lines) as a function of energy of levels given for the vector $J=0 \to 2$. The red and blue lines correspond to lower and upper members of the coupled states ($J^+_1$ and $J^+_2$).
The open circles (open squares) show the results of our calculations for the states $J^+_1$ (red) ($J^+_2$, (blue)), $J=0,2$ (see Fig. \ref{fig:7}).
The energy is measured from the three-body $\alpha+\alpha+\alpha$ threshold.
} \label{fig:8}
\end{figure}

Comparing the effective mass for $0^+_1$ and $2^+_1$ states,
one may detect an energy dependence or an orbital momentum dependence of the effective mass.
This comparison is illustrated by Fig. \ref{fig:8}, where we show the effective mass ($m^*/m$) variations for the $J=0 \to 2$ vector using data from Tabl. \ref{t8}. The mass variation can be represented by the form:
$ \Delta m^*=({\delta}m^*/{\delta}J)\Delta J+({\delta}m^*/{\delta}E)\Delta E$ assuming a dependence of effective mass on orbital momentum $J$ and energy $E$.
One can show that $({\delta}m^*/{\delta}J)=0$ and the effective mass depends on the energy only. The variations of the effective mass for different energies when the orbital momentum is fixed $J^+_1$ ($J^+_2$), $J=0,2$ (see Fig.~\ref{fig:7}) can be described by the form: $ \Delta m^*=({\delta}m^*/{\delta}E)\Delta E$. In Fig.~\ref{fig:8},
the corresponding results are shown by open circles and squares. The results obtained for both representations coincide. Thus, the effective mass does not have essential orbital momentum dependence for $0^+_n$ and $2^+_n$ states when $n=1,2$.
In addition, it can be seen that the dependence of the effective mass on energy can be represented as a linear function.
For states $J^+_n$, where $J=0.2$ and $n=1.2$, the slopes of the functions are approximately equal.
At the same time, the linear functions are different in constant for the states $n=1$ and $n=2$.

\begin{figure}[t]
\begin{center}
\includegraphics[width=19pc]{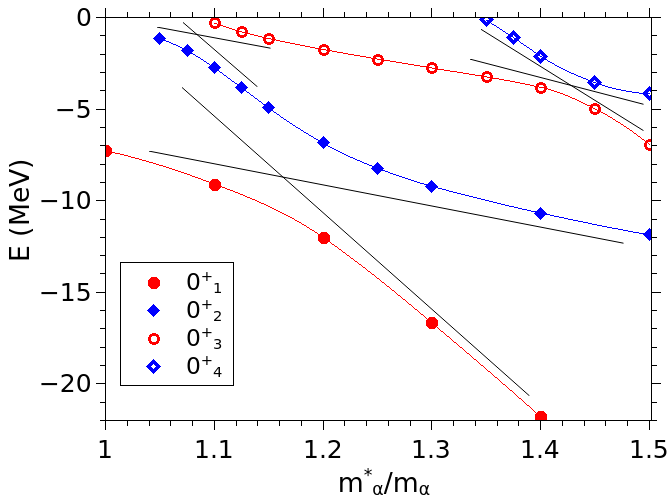}
\includegraphics[width=19pc]{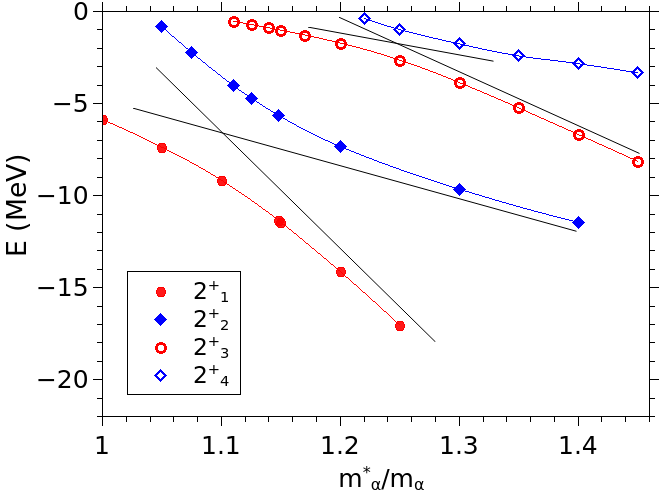}
\end{center}
\caption{
The energy $E$ of the low lying levels $J^+_1$ (red solid and open circles) and $J^+_2$ (blue solid and open squares), $J=0,2$ of the $3\alpha$-cluster system along of the $\alpha$-particle effective mass $m_\alpha^*/m_\alpha$. The Upper panel (Lower panel) corresponds to
$0^+$ ($2^+$ ) band. The assumed level crossing is schematically shown by intersecting thin lines for the case when there is no coupling between the levels.
The three-body potential Eq. (\ref{v_3}) was added to the model. The energy is measured from the $\alpha+\alpha+\alpha$ threshold.
} \label{fig:77}
\end{figure}

The results shown in Fig. \ref{fig:7} can be interpreted in the terminology of the two-level theory as anti-crossings of levels \cite{CT}.
The $n=1$ and $n=2$ states are the lower and upper states of a quasi-doublet formed by the anti-crossed levels.
To clarify this assumption, we performed calculations with the three-body potential (\ref{v_3}).
This potential does not affect the two-body $\alpha-\alpha$ threshold, and we can make an extension for allowable energy and effective-mass scopes to apply the numerical procedure proposed above for bound states. The results of the calculation are given in Fig. \ref{fig:77}. The three-body potential increases the depth of the potential well where are the bound states $J^+_1$, $J^+_2$, $J^+_3$, $J^+_4$, $J=0,2$.
One can see that the anti-crossing of the levels occurs for the pairs $0^+_1$ and $0^+_2$, $0^+_2$ and $0^+_3$, $0^+_3$ and $0^+_4$. In the $2^+$ states, the similar situation is visible for the pairs $2^+_1$ and $2^+_2$, $2^+_3$ and $2^+_4$. According to the two-level system theory, we can definite upper member and lower member of the corresponding quasi-doublets.
In the absence of a coupling between levels, the levels may cross. The assumed crossing of the levels is schematically presented by crossing fine lines in Fig. \ref{fig:77}.
One can assume that the level anti-crossings will be saved in the continuous spectrum of the nucleus.
Thus, our study is extended to the states of the continuous spectrum.
\begin{figure}[!t]
\begin{center}
\includegraphics[width=19.5pc]{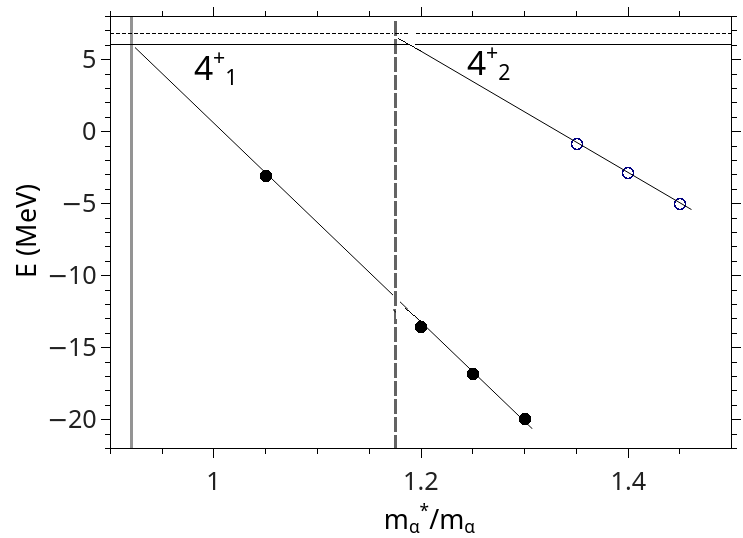}
\includegraphics[width=18.5pc]{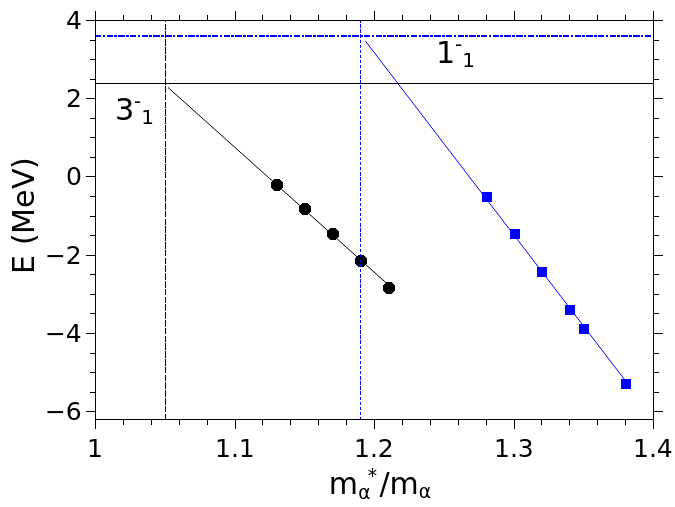}
\end{center}
\caption{
The energy $E$ of the $^{12}$C low lying levels and the $\alpha$-particle effective mass $m_\alpha^*/m_\alpha$.
(Upper panel) The results for the levels $4^+_1$ and $4^+_2$ (solid and open circles) are given. (Lower panel) The results for $3^-_1$ (solid circles) and $1^-_1$ (solid squares) states are presented. The experimental data are shown by the solid/dashed horizontal lines. The vertical lines represent corresponding values of the effective masses.
The fine solid lines are given for eyes to show a linear extrapolation.
The energy is measured from a three-body $\alpha+\alpha+\alpha$ threshold.
} \label{fig:9}
\end{figure}

We can conclude that there is tunneling coupling between the $0^+$ levels and between the $2^+$ levels. 
At the same time, the levels with different $J$ are not coupled due to the different rotation symmetry of these states. The coupling parameter depends on overlapping wave functions\cite{FMV,FMV1} and is small for these states.
The $4^+_1$ and $4^+_2$ levels do not demonstrate also a coupling. The results of our calculations for the levels $4^+_1$ and $4^+_2$ are given in Fig. \ref{fig:9}(Upper panel).
The effective masses for these levels have a significant difference, 0.92$m_\alpha$ and 1.17$m_\alpha$. That gives different contributions of the kinetic energy to the Hamiltonians and provides the difference in coordinate behavior of the wave functions. One can assume that the last difference arises from the effect of coordinate scaling due to the mass difference. The asymptotic behavior of these functions differs only in the mass coefficient due to the closeness of the energies of the levels.

The energy/effective mass dependence is a simple linear function with different slopes for the $4^+_1$ and $4^+_2$ levels. One can interpret that the corresponding three-body potential must have different depths in the $4^+_1$ and $4^+_2$ states to reproduce the experimental data for these levels. Thus, the results of \cite{H13}, in which the three-body potential acts with the same depth for different $4^+$ levels, can be explained within the framework of the effective-mass approach.

The results for the states $3^-_1$ and $1^-_1$ are shown in Fig. \ref{fig:9}(Lower panel). Also, a linear dependence of energy on the effective mass has been found for these states.
There is no coupling between the levels due to different rotational symmetries and significant differences in the effective mass.

\section{Conclusions}
The imperfectness of $\alpha$-cluster structure of the $^{12}$C nucleus
can be described with a three-body force. Based on the effect of mass-energy compensation, the three-body force can be taken into account in a phenomenological manner using the effective-mass approach.

We have presented the results of the Faddeev calculations for the bound 3$N$ system to demonstrate
the mechanism of mass-energy compensation for a three-body Hamiltonian. Based on this compensation effect, we have proposed the effective-mass approach which was initially applied for a bound few-body system.
In the presented work, this approach is assumed to be applicable to describe the resonances states.
Based on this assumption, we have defined the effective mass of the $\alpha$-cluster in the $^{12}$C. The corresponding mass adjustment was performed to reproduce the experimental data for several low-lying states available with the Ali-Bodmer $\alpha$-$\alpha$ potential.

For the $0^+$ (and $2^+$) band, we have found that there is a coupling between pairs of nearest levels.
We describe these coupled levels as quasi-doublets following the terminology of the theory of two-level systems \cite{CT}. The anti-crossings picture, which was obtained on the energy/effective-mass plane, is interpreted within this theory.
According to this interpretation, the coupling matrix elements of the quasi-doublets are non-zero due to overlapping the wave functions of the nearest levels. The last one is possible, for example, when these levels have the same orbital momentum and similar rotation symmetry.
We have found that the coupling takes place for the pairs of the levels $0^+_1$ and $0^+_2$, $0^+_2$ and $0^+_3$, $0^+_3$ and $0^+_4$. Also, we have found that the structure of the $2^+$ band repeats the one for the $0^+$ band.
Along with that, the levels with different orbital momenta like the $0^+$ and $2^+$ levels are not coupled.
However, the levels $4^+_1$ and $4^+_2$ are not coupled also despite the same orbital momenta. Here, the orbital momentum similarity of wave functions is broken by differences in the kinetic term in the Hamiltonian due to the significant difference in the effective masses.
The similar explanation one can give for $3^-_1$ and $1^-_1$ resonances. There is no coupling due to different rotation symmetry and significant differences in the effective masses.

Within our approach, an effective-mass value is correlated with the strength of a three-body potential. However, this approach does not support the assumption about the direct dependence of the effective mass (and 3BF) on only the total orbital momentum used previously within the approximations \cite{FSV2005,H13} for three-body potentials.
We conclude that 3BF  has to be separately defined for each energy level as it was earlier concluded in Ref. \cite{Fedorov}. 
Note that, the effective mass of the $0^+_1$ ground state is the largest in the considered 3$\alpha$ spectrum.
In other states, the mass of the $\alpha$-cluster increases or decreases relative to the value of free mass.

\section*{Acknowledgments}
This work is supported by the National Science Foundation Centers of Research Excellence in Science and Technology NSF CREST 1647022 and,  the Department of Energy/National Nuclear Security Administration award NA0003979,  NSF EiR award number DMR-2101220.


\end{document}